\documentclass[12pt]{article}

\usepackage{latexsym} %for \Box
\usepackage{mathrsfs} %for \mathscr
\usepackage{amsthm} %for \theoremstyle{definition}

\newcommand{\mr}{\mathscr R}
\newcommand{\mi}{\mathscr I}

\begin{document}

\pagestyle{myheadings}  \markright{	 Eric Werner, A Category Theory of Communication Theory}

\title{A Category Theory of Communication Theory}
\author{Eric Werner \thanks{Balliol Graduate Centre, Oxford Advanced Research Foundation (http://oarf.org).
\copyright Eric Werner 2015.  All rights reserved. }\\ \\
University of Oxford\\
Department of Physiology, Anatomy and Genetics, \\
and Department of Computer Science, \\
Le Gros Clark Building, 
South Parks Road, 
Oxford OX1 3QX  \\
email:  eric.werner@dpag.ox.ac.uk\\
%Website: http://ericwerner.com
}

\date{}
\maketitle

\begin{abstract}
A theory of how agents can come to understand a language is presented.  If understanding a sentence $\alpha$ is to associate an operator with $\alpha$ that transforms the representational state of the agent as intended by the sender, then coming to know a language involves coming to know the operators that correspond to the meaning of any sentence. This involves a higher order operator that operates on the possible transformations that operate on the representational capacity of the agent.  We formalize these constructs using concepts and  diagrams analogous to category theory. 
\end{abstract}

\pagebreak

\section{Communication as operators on agent representations}

Let $\mi$ be the interpretation of a language $L$. $\mi$ maps signals $\alpha \in L$ to operators $\mi (\alpha)$.  An operator $\mi(\alpha)$ transforms representations $\mr$ into new representations $\mi(\alpha)(\mr) = \mr^{\alpha}$. Hence, $\mi(\alpha)$ is a transformation on transformations, or what is known as a functor in category theory. Transformations are designated by arrows.  A given interpretation $\mi$ of a language $L$ shows how each sentence as message or more generally signal $\alpha$ of the language transforms the representational state $\mr$ of a receiver.  In learning a language, an agent starts in a state of total ignorance about the interpretation $\mi$ and gradually through interaction with its social environment of other agents gains more and more information about the interpretation itself.  Thus, the learning of language involves yet another level of transformation, namely, it transforms the information about $\mi$.  If $\mi^\Omega$ represents total ignorance about the actual interpretation $\mi$ of the language $L$ (The interpretation is relative to a society since different societies could interpret the same language differently) and $\mi^{PI}$ represents perfect information, then the process of coming to understand a language involves a series of transformations that take the agent from total non-understanding $\mi^\Omega$ to perfect understanding $\mi^{PI}$ or some state $\mi^X$ in between. 

Viewed extensionally or set-theoretically (see \cite{Khinchin49, Shannon48, vonNeumann47, Werner88a, Werner88b, Werner89}) the state of uncertain about the interpretation $\mi^X$ is a set of possible interpretations $\mi^1, \dots , \mi^{k(X)}$.  Viewed positively 
$\mi^X$ is a partial representation of the interpretation, partial both in terms of the domain it covers as well as the values of the mapping.  So, for example, $\mi^X(\alpha)$ may not map to a unique representational operator and, instead only pick out a set of operators, or viewed positively, a partial operator on agent representations\footnote{The relationships to operators in quantum mechanics \cite{vonNeumann55, Everett57} and statistical mechanics \cite{Khinchin49} are noted}. 

Coming to learn a language can then be viewed as a path or history
 
$$H^\mi = \mi^\Omega = \mi^{X_{t_1}} \rightarrow \mi^{X_{t_2}} \rightarrow  \dots \rightarrow \mi^{X_{t_z-1}} \rightarrow \mi^{X_{t_z}}  = \mi^{PI}$$

Each such a path $H^\mi$ is a possible path to coming to understanding a language. Let $\psi^\mi$ be the set of all possible paths to learning the interpretation $\mi$ of a language $L$.  Each transformation step in 
$H^\mi$ is associated with an interaction between the agent $A$ and the social environment $M$ such that 

$$(\mr_A \otimes \mr_M)_{t_1} \dots (\mr_A \otimes \mr_M)_{t_n}(\mi^\Omega) = \mi^{X}_{t_n} $$

In other words, the interactions on $\mi^X$ progressively reduce the uncertainty the agent $A$ has about the interpretation $\mi$ of the language $L$.  This means these interactions presuppose a meta operator that acts on the representation an agent $A$ has of the interpretation $\mi^X$.  The meta operator thus maps the possible interpretations (allowed by the agent's available meta-interpretation information) to a new set of possible interpretations.

$$A \otimes M: {\mi^\Omega}^2 \rightarrow   {\mi^\Omega}^2$$

 where ${\mi^\Omega}^2$ is the power set (set of all subsets of $\mi^\Omega$ ).
 
 This means that the agent has to have a meta-representational operator capacity that enables it to make the meta-representational transformations necessary for learning the meaning of a language. Granted, the implementation of such meta-transformations in humans may be messy, complex  and downright incomprehensible having themselves been formed by the vagaries and randomness the ultimate meta-transform-evolution. Still, no matter how complex,  they are meta-transformations and as such can be studied from the abstract perspective of set theory and category theory.  

\section{The meta space of all possible interpretations} 

\begin{equation}
\mi^\Omega = \left\lbrace
\begin{array}{c}

\mi^1 = \left\{
\begin{array}{c}
\mi^1(\alpha_1): \mr \Rightarrow \alpha^{\mi^1}_1(\mr) \\ \\
\vdots \\ \\
\mi^1(\alpha_n): \mr \Rightarrow \alpha^{\mi^1}_n(\mr) \\
\end{array} 
\right. \\
\\
\vdots
\\ \\
\mi^k = \left\{
\begin{array}{c}
\mi^k(\alpha_1): \mr \Rightarrow \alpha^{\mi^k}_1(\mr) \\ \\
\vdots \\ \\
\mi^k(\alpha_n): \mr \Rightarrow \alpha^{\mi^k}_n(\mr) \\
\end{array}
\right.

\end{array}
\right.
\end{equation}
\\

For any $\mi^i \in \mi^\Omega$ the interpretation maps $\mi^i$ signals $\alpha \in L$ to operators on $\mr^\Omega$. Where $\mr^\Omega$ is the set of all possible representational states of an agent. We call $\mr^\Omega$ the {\em representational capacity} of the agent. Hence, $$\mi^i :  \mr^\Omega \Rightarrow \mr^\Omega$$

$$\psi^\Omega = \{X : X \subset \mi^\Omega \} = {\mi^\Omega}^2$$ is the power set of $\mi^\Omega$ and is the set of all possible meta-information states about the interpretation $\mi$.
\\
%\begin{table}
%\caption{The meta-interpretation space}
%\end{table}

\pagebreak

\end{document}